# Magnetic phase transitions in plasmas and transport barriers

Emilia R. Solano[1] and Richard D. Hazeltine[2]

[1]Asociación EURATOM-CIEMAT para Fusión, Av. Complutense, 22, E-28040, Madrid, Spain

[2]Institute for Fusion Studies and Department of Physics, The University of Texas at Austin, Austin, TX 78712, USA

E-mail: emilia.solano@ciemat.es

A model of magnetic phase transitions in plasmas is presented: plasma elements with pressure excess or defect are dia- or para-magnets and move radially under the influence of the background plasma magnetisation. It is found that magnetic phase separation could be the underlying mechanism of L to H transitions and drive transport barrier formation. Magnetic phase separation and associated pedestal build up, as described here, can be explained by the well known interchange mechanism, now reinterpreted as a magnetisation interchange. The interchange mechanism can drive motion of plasma elements even when stable. A testable necessary criterion for the L to H transition is presented.

## 1. Introduction:

Plasma confinement transitions and the spontaneous creation of transport barriers in magnetically confined plasmas have been extensively studied since their discovery [1]. It is conventionally believed that the high confinement regime is achieved when the velocity $\vec{E} \times \vec{B}$ shear is sufficient to stabilise plasma electrostatic fluctuations responsible for anomalously high transport [2,3,4,5,6,7,8]. Nevertheless the transition trigger mechanism remains elusive [9].

Despite the overwhelming body of work searching for an electrostatic mechanism for the L to H transition (see references above), over the years there have been some indications that plasma diamagnetism plays a role in the formation of transport barriers [10,11,12,13,14]. Additionally, the spectacular MAST movie [15] of an L to H transition inspired our study of phase transitions in plasmas.

Here we present a study of the dynamics of magnetic phase transitions in plasmas, and show how they influence plasma confinement, leading to phenomenology akin to the experimental observations.

A clarification before we proceed: this is not a conventional study of linear MHD stability, nor of transport driven by electrostatic fluctuations. We present a fundamentally novel approach to the problem.

The mechanism we propose is very simple: low pressure plasmas are paramagnetic and attract low pressure plasmas, becoming more paramagnetic. Conversely, high pressure plasmas are diamagnetic and attract diamagnets. Therefore the magnetisation state of the background plasma, para or diamagnetic, determines the motion of higher or lower pressure plasma elements, driving magnetic phase separation and pressure profile segregation in regions with high and low gradients.

This paper is organised as follows: in section 2 we describe plasma magnetisation and the motion of magnetised plasma elements under the influence of background plasma magnetisation. In section 3 we discuss phase separation. In section 4 we show how blob convective motion affects profile evolution. In section 5 we connect to interchange theory to obtain some estimates on blob behaviour, while on Section 6 we discuss potential experimental tests of the model.

## 2. Global and local plasma magnetisation

Magnetisation is the reaction of a material to an externally applied magnetic field. In tokamaks the dominant extrinsic magnetic field is toroidal. Poloidal currents flowing in the plasma and enclosing the magnetic axis can be diamagnetic, decreasing this externally applied toroidal field, or paramagnetic, increasing it. The sign of the poloidal current characterises the magnetisation state of a given plasma region. The boundary between para and diamagnetic radial regions has zero magnetisation: $j_\theta = 0$ .

We begin to illustrate these concepts by revisiting plasma equilibrium as described in conventional textbooks [16,17]. Consider a plasma cylinder, with an externally imposed longitudinal magnetic field $B_z$ and a pressure profile peak at the centre, as depicted in red at the top of Fig. 1a. Assume equilibrium

$$\boldsymbol{F} = n\,m\,\frac{d\boldsymbol{v}}{dt} = -\nabla p + \boldsymbol{j} \times \boldsymbol{B} = \boldsymbol{0} \tag{1}$$

Taking the perpendicular component of (1) we have the perpendicular current $\mathbf{j}_\perp = \mathbf{b} \times \nabla p / B$ . It is often referred to as the diamagnetic current because its poloidal component reduces the externally applied $B_z$ field. Conversely, a plasma pressure hollow (blue in Fig. 1a, at the bottom) would be surrounded by a perpendicular paramagnetic current whose poloidal component would amplify the externally applied field.

If there is a longitudinal free current $\mathbf{j}_z$ flowing in the plasma, the magnetic field $\mathbf{B}$ becomes helical. A current parallel to $\mathbf{B}$, $\mathbf{j}_\parallel$, will now have a poloidal component. As illustrated in Fig. 1b, if $\mathbf{j}_\parallel \cdot \mathbf{j}_z > 0$, then its poloidal component, $\mathbf{j}_{\parallel\theta}$, adds to the background longitudinal field: it is paramagnetic. For instance, the poloidal component of the bootstrap current in tokamaks is paramagnetic.

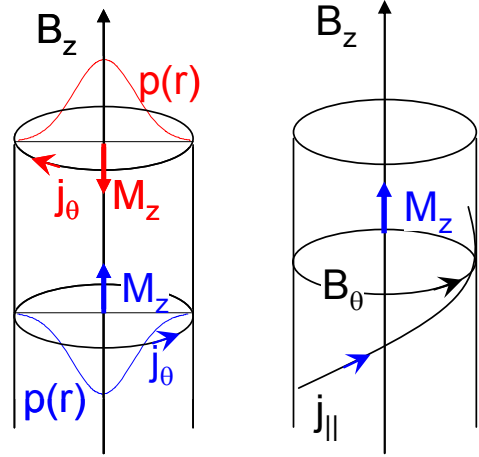

The magnetic phase of a plasma radial region is given by the sign of the poloidal current density flowing in that region: diamagnetic (paramagnetic) when it reduces (increases) the external field inside of that radius.

Fig. 1a: sketch of plasma magnetization associated with pressure gradients in cylindrical plasma. 1b: magnetisation associated with co-parallel current.

Obviously, the poloidal field produced by a free current flowing in the plasma is not considered as magnetisation.

In a torus the geometry is more complicated, but these basic concepts of dia and paramagnetism remain: perpendicular current around a pressure hill leads to diamagnetism, while perpendicular current around a pressure hole and co-parallel current leads to paramagnetism. In a toroidally symmetric system with sub-sonic rotation the background poloidal current density $j_\theta$ and the toroidal flux density F are given by

$$j_\theta = -F'B_\theta / \mu_0, \text{ and } F(\Psi) = RB_\varsigma \tag{2}$$

where $B_\zeta$ is the toroidal magnetic field. Note that F is a flux function, constant on flux surfaces labelled by poloidal flux $\Psi$, and $F' \equiv dF/d\Psi$. Therefore background plasma phase, given by the sign of $j_0$, is well defined in radial regions of a toroidal tokamak, as it is in a cylinder.

To simplify matters we will carry on the explanation in a "cylindrical tokamak".

Integrating (1) in a cylindrical tokamak results in a simple relationship between plasma longitudinal magnetization and plasma pressure, p [16]:

$$\beta_\theta - 1 \equiv \frac{\int_0^a p \, dS}{\pi a^2 B_{\theta a}^2 / 2\mu_0} - 1 = \frac{B_{za}^2 - \langle B_z^2 \rangle}{B_{\theta a}^2} \simeq \frac{2 B_{za} \left( B_{za} - \langle B_z \rangle \right)}{B_{\theta a}^2} \tag{3}$$

Here the angle brackets indicate a cross-sectional average, a is the plasma radius, $B_{za}$ the externally applied longitudinal field at the boundary r=a, $B_{0a}$ is the poloidal field at the boundary. $\beta_\theta - 1$ measures the balance between para and diamagnetism, and therefore the total toroidal magnetization of the plasma. A global magnetic phase transition of the plasma can take place if central heating increases the pressure sufficiently to raise $\beta_\theta$ above 1.

So far we have discussed background plasma magnetisation. Next we describe the magnetisation of a plasma fluid element embedded in the background plasma. We begin by assuming that pressure fluctuations are present in the plasma. Consider the background plasma in equilibrium, with pressure $p_0$, confined by a magnetic field $\mathbf{B}$. Consider next a small plasma element that for some reason has higher pressure than the background. Its characteristic length along the field is much longer than its transverse radius: $\lambda_\parallel \gg \lambda_\rho$. Such plasma structures are often described in the literature, sometimes named Intermittent Plasma Objects [18], most times they are unglamorously named blobs [19]. At the blob location the total pressure is $p = p_0 + \tilde{p}$, tildes describing blob properties. The centre of the blob is located at the field line where $\nabla(\tilde{p}) = 0$, and its radius is determined by the nearest location where $\nabla(p_0 + \tilde{p}) = 0$.

The perpendicular current density associated with the blob pressure profile is

$$\tilde{\mathbf{j}}_\perp = \frac{\mathbf{b} \times \nabla \tilde{p}(\rho)}{\bar{B}} = \nabla \times \tilde{\mathbf{M}} \tag{4}$$

Integration of (4) results in the local magnetisation [20- Chapter 4.5]

$$\tilde{\mathbf{M}} = \frac{1}{\lambda_\parallel} \int_0^\rho \frac{\mathbf{b}}{B} \frac{\partial \tilde{p}(\rho')}{\partial \rho'} \lambda_\parallel \, d\rho' \approx -\frac{\tilde{p}}{B} \mathbf{b} \tag{5}$$

The negative sign indicates that a pressure hill is locally diamagnetic, while a pressure hole is paramagnetic. Blob magnetisation is illustrated in Fig. 2 for both types of blobs, hills and holes. (Note: for now we assume that there is no fluctuation in the parallel current density inside the blob).

The radial force balance equation describing the trajectories of the magnetised plasma blobs under the influence of the background plasma magnetization is obtained by Taylor expansion of the magnetic field from the centre of mass of the tube and integrating over the blob volume, as described in electromagnetism textbooks [21]:

$$m_{blob} \frac{d\mathbf{v}_{blob}}{dt} = \int_V \left( -\nabla \tilde{p} + \tilde{\mathbf{j}} \times \mathbf{B} \right) dV = \int_{\tilde{V}} \left( \left( \overline{-\nabla \tilde{p} + \tilde{\mathbf{j}}_\perp \times \mathbf{B}_0} \right) + \tilde{\mathbf{j}}_\perp \times (\boldsymbol{\rho} . \nabla \mathbf{B}) \right) dV \tag{6}$$

The 1$^{st}$ parenthesis in (6) is zero because the perturbed diamagnetic current density exactly balances the perturbed pressure (this property of blobs is commonly called flute cancellation). Carrying out the integral and dividing by the blob volume, $\tilde{V}$, we obtain the magnetization force density of the background plasma acting on the para or dia-magnetic plasma blobs, which concerns us in this study:

$$\overline{\rho}_{m,blob} \frac{d\mathbf{v}_{blob.}}{dt} = \frac{1}{\overline{V}} \int_{\overline{V}} \left( \nabla \left( \tilde{\mathbf{M}} \cdot \mathbf{B} \right) \right) dV \tag{7}$$

From now on we ignore any possible poloidal component of blob magnetisation and concentrate on the toroidal magnetisation $\tilde{M}_\zeta$. Further we assume for simplicity that the background plasma magnetisation exceeds the radial gradient of the blob's magnetisation, $\tilde{M}_\zeta \nabla B_\zeta >> B_\zeta \nabla \tilde{M}_\zeta$. The toroidal magnetization force is especially easy to understand in the cylindrical tokamak:

$$\overline{\rho}_{m,blob} \frac{dv_r}{dt} = \tilde{M}_\zeta \frac{d\overline{B}_\zeta}{dr} \tag{8a}$$

$$= -\mu_0 \tilde{M}_\zeta \overline{j}_\theta \tag{8b}$$

The overbar indicates an average of the background quantity over the blob volume, $\overline{\rho}_{m,blob}$ is the average blob mass density. Equation (8a) shows that the toroidal magnetic field of the background plasma provides an anti-potential for motion of local magnetized blobs. In a paramagnetic region $d\overline{B}_\zeta / dr < 0$ and a paramagnetic blob is driven inward, up the hill of the paramagnetic field. We will return to this in section 4.

### 3. Phase separation, heuristic:

In the cylindrical case consider the behaviour of a para or dia-magnetic blob in the presence of a longitudinal magnetic field with a radial gradient generated by poloidal currents in the background plasma, equations 8a,b. Diamagnetic hot blobs move down the magnetic hill, taking heat away from the high field region. Paramagnetic cold blobs move up the magnetic hill taking cold plasma to the high field region. If there is a heat source at the plasma core and a heat sink at the outer edge, and the plasma is paramagnetic everywhere, the exchange of blobs contributes to overall outward transport of pressure, in exchange for inward transport of toroidal paramagnetism. In terms of an energy principle, a decrease in plasma kinetic pressure is compensated by an increase in magnetic pressure. This tendency towards increasing paramagnetism might explain L-mode (low confinement). If there is a diamagnetic region, it attracts hot blobs and expels cold ones: increasing diamagnetism would explain H-mode (localised high confinement).

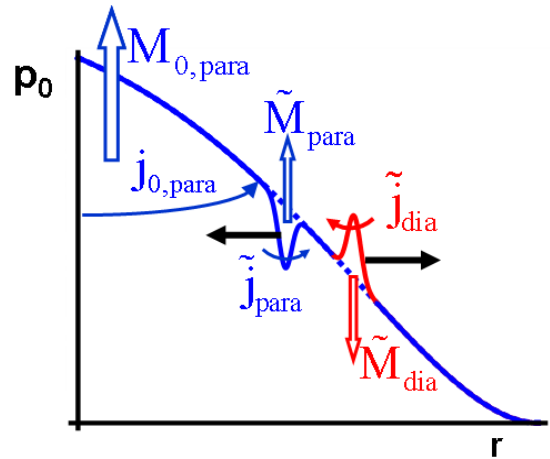

Fig. 2 Paramagnetic background plasma pressure profile with high (red) and low (blue) pressure blobs, individually magnetised. Bulk poloidal current density and associated magnetisation are denoted with a subscript "0", while local quantities associated with blobs are denoted with a tilde. Black arrows indicate magnetisation force direction.

A sketch of the plasma pressure profile with some pressure hills and holes is shown in Fig. 2, indicating how positive (extra pressure) and negative (pressure defect) blobs would move under the influence of the plasma magnetisation force (8).

Consider Fig. 2: with typical monotonically decreasing background pressure profiles, the mechanism of phase separation looks like a growing instability in paramagnetic regions: as a cold blob moves inwards towards higher pressure its "amplitude" (the difference between $\tilde{p}$ in the blob centre and the ambient pressure) increases. A hot blob moving outwards also would appear to grow as it moves outwards towards lower pressure regions. On the other hand, if the centre of the plasma is diamagnetic a hot blob moves inwards up the pressure gradient until it eventually encounters a matching background pressure. The blob then merges with the background pressure (adding to its gradient) and ceases to exist. Through this process the "amplitude" of the hot blob decreases, so we would classify this behaviour as stable, self-limiting.

We have therefore a magnetic phase transition taking place because of blob convection: initially there can be diamagnetic domains (the blobs), surrounded by a paramagnetic background. As blobs move and alter the magnetic field, phase separation takes place and a poloidally complete diamagnetic layer can form around a central paramagnetic volume.

### 4. Pedestal build-up:

Let us consider the time evolution equation of the background pressure, where divergence of heat flux is driven by heat sources and sinks:

$$\frac{3}{2}\frac{\partial p}{\partial t} + \nabla \cdot \mathbf{Q} = H \tag{9}$$

and for the magnetic field,

$$\frac{\partial \mathbf{B}}{\partial t} - \nabla \times (\mathbf{v} \times \mathbf{B}) + \nabla \times \eta \left( \mathbf{j} - \mathbf{j}_{ni} \right) = 0 \tag{10}$$

where non-inductive current drive and bootstrap (both non-inductive) do not contribute to Ohm's law. The contribution of the magnetization force (subscript M) to evolution of surface averaged (<...>) pressure changes and toroidal magnetic field change can be estimated from the convective flux of blobs as

$$\left\langle \frac{3}{2}\frac{\partial p}{\partial t} \right\rangle_M = -\nabla \cdot \left\langle \tilde{p} \, \mathbf{v}_r \right\rangle \tag{11}$$

$$\left\langle \frac{\partial \mathbf{B}_z}{\partial t} \right\rangle_M = \nabla \times \left\langle (v_r B_z \boldsymbol{\theta}) \right\rangle \tag{12}$$

In (11,12) we have only written explicitly the contribution from the magnetisation force (8).

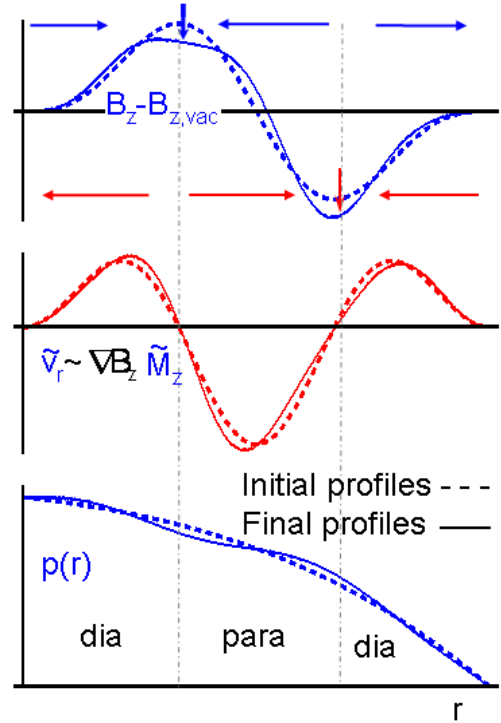

Fig. 3: Plasma evolution cartoon (initial states in dashed lines). a) toroidal field in magnetized plasma, b) velocity of paramagnetic blobs and c) radial pressure profile.

Additional terms would describe the effect of particle, momentum and heat sources and sinks, electrostatic fluctuations, collisional transport, resistivity and non-inductive current drive. These numerous competing effects are not discussed here, we pursue the simplest model for illustration purposes.

With an initial model pressure profile depicted in Fig. 3.c (dashed), and an assumed toroidal magnetic field profile, shown in Fig. 3.a (dashed), we evaluate the effect of a radial velocity proportional to the magnetisation force, Eq. (8). The velocity of paramagnetic tubes is shown in Fig. 3.b, and represented by horizontal blue arrows in Fig. 3a. Paramagnetic plasma elements seek high magnetic field, while diamagnetic elements take their heat to the magnetic well (red arrows). The changed magnetisation and pressure are illustrated with solid lines in 3.a, 3.c. We see that the magnetisation force naturally creates a pedestal structure by adding heat to the magnetic well, and removing it from the magnetic hill: pressure is flattened at the pedestal top, steeper outwards, forming a "transport barrier". Further, the tendency is for the well to become narrower and deeper, while the hill becomes broader. Fig 3 is only a cartoon: it does not attempt to model other transport mechanisms, or the effect of central heating sources and edge heat sinks. Note that pedestal build-up associated with plasma diamagnetism has been described earlier [11], with a rather different model (resistive ballooning modes vs. drift waves). Their model is numerically detailed, but lacks the generality of ours.

## 5. Magnetisation interchange:

The magnetisation force given by equation (8) is a radial force that acts equally on both ions and electrons, and therefore can give rise to interchange instabilities, the plasma equivalent of the Raleigh-Taylor instability. The transport mechanism associated with the magnetisation force can be most easily understood as a magnetisation interchange (not linear) when blobs have been formed and drift under the influence of magnetic field gradients. Plasma thermal pressure and toroidal magnetic field pressure are interchanged by blob motion. The interchange instability is explained in detail in [22], using gravity as the radial force. Here the plasma background magnetisation replaces gravity, with the peculiarity that it can have either sign.

Whether seed fluctuations are due to magnetisation interchange instabilities or not, magnetisation interchange would drive blob radial movement and affect the evolution of the pressure profile at magnetisation phase boundaries. We must recall that curvature-driven interchange of blobs is believed to govern transport in open field lines [19,23], as shown for instance in beautiful simulations by O.E. Garcia [24]. Even the radial motion of hot and cold blobs in different directions (away from bulk plasma or towards it) has been characterised experimentally [18]. By contrast the magnetisation interchange we present here takes place in the closed field line region.

The conventional formalism of the gravitational interchange instability provides estimates for interchange linear growth rates [22], which we can use to estimate growth rates for the magnetisation interchange

$$\gamma = \sqrt{g \kappa_\perp} \simeq \sqrt{\frac{\tilde{M}_\zeta}{\bar{\rho}_{m,blob}} \frac{d\bar{B}_\zeta}{dr} \frac{1}{\lambda_\rho}} \simeq \sqrt{-\frac{1}{\bar{\rho}_{m,blob}} \frac{\tilde{p}}{\bar{B}} \frac{d\bar{B}_\zeta}{dr} \frac{1}{\lambda_\rho}} , \qquad (13)$$

Here we have mapped the gravitational force, $\mathbf{g}$, acting on the blob with mass density $\bar{\rho}_{m,blob}$, with the magnetization force acting on the magnetic moment $\tilde{M}_\zeta$ (equations (8)), and we have associated the blob transverse size $\lambda_\rho$ to the inverse wavelength, $\kappa_\perp \sim 1/\lambda_\rho$. An estimate of the radial velocity of the blobs is given by $\lambda_\rho \gamma$. Inspection of the signs in equation (13) show a positive growth rate for mixed states: hot diamagnetic blobs in a paramagnetic plasma, and cold paramagnetic blobs in a diamagnetic plasma. But as we mentioned earlier, growth rate does not tell the whole story. The seed pressure fluctuations may or may not originate in magnetization interchange instabilities. If they do not, the creation rate of blobs would control transport rates, rather than the growth rate of the blobs themselves.

Nevertheless, we see that the linear growth rate of the instability is inversely proportional to B, facilitating a phase transition at low field. Low background density, atomic mass and impurity content would imply "lighter" blobs (low mass density), also facilitating the transition. These tendencies coincide with experimental observations.

Both magnetic and rotation shear provide stabilization for cylindrical curvature driven interchange instabilities, opposing the interchange mechanism. Linear stability of ideal interchange modes in the presence of magnetic shear in a cylindrical tokamak (due to energy required for line bending) has long been known as the Suydam criterion [25, 20-Chapter 7]. It can be written in terms of the magnetisation force on a magnetised layer (no longer a localised blob) with magnetisation $\bar{M}_z$ as:

$$-\frac{1}{B^2/2\mu_0} \frac{dp}{dr} \frac{1}{B^2/\mu_0} \frac{d}{dr}\left(\bar{M}_z \bar{B}_z\right) \simeq -\frac{dp/dr}{B^2/2\mu_0} \frac{1}{B^2/\mu_0}\left[\bar{M}_z \frac{d\bar{B}_z}{dr}\right] > \frac{q'^2}{4q^2} \quad \text{(for instability) (14)}$$

Here we see again that mixed states (diamagnetic layer in a decreasing $B_z$ region or vice-versa) enhance the instability. We also obtain a criterion for the minimum shear required to stabilise the magnetization interchange. And we see that minimising magnetic shear is beneficial but not essential for magnetic phase separation and therefore the creation of transport barriers, in agreement with MHD predictions [26,27,28] and experimental evidence [29, 30, 31, 32, 33].

In toroidal geometry magnetized blobs with short $\lambda_\parallel$ can be affected by the 1/R variation of the external toroidal field. In the low field side (LFS) short hot blobs are driven out of the plasma, while in the high field side (HFS) they are driven towards the plasma core. Only blobs that average out the field gradient of the externally imposed toroidal field ($\lambda_\parallel > qR$) will be sensitive to background plasma toroidal magnetization. Further, the plasma temperature in the blobs must be high enough to allow the plasma particles to sample LFS and HFS (sufficiently low collisionality, or equivalently, sufficiently low density). These pre-conditions may relate to the $n_e B_\zeta$ scaling of the power threshold for the L to H transition [34].

**6. Proposed experimental tests:**

Let us now consider possible tests of this model in the case of an L to H transition in a tokamak. In L-mode, while the plasma pressure is sufficiently low, the plasma is paramagnetic and attracts cold blobs: the magnetization instability cools the plasma and reduces confinement. As further central heating is applied the plasma pressure increases, increasing the pressure gradient and diamagnetism, partly compensating the toroidal paramagnetism produced by the parallel current. Eventually a critical pressure gradient is reached, creating a diamagnetic region in the plasma, where magnetic shear is overcome and a pedestal can develop. The movement of para- and diamagnetic blobs, impelled by the magnetisation force, drives phase separation and builds up the pedestal. In the pedestal gradient region confinement is improved.

As an aside we would note that the pedestal build up automatically affects momentum balance and can lead to changes in $E_r$. In the presence of a monotonically decreasing density profile the high pressure blobs that move inward into a diamagnetic layer bring lower density, while cold blobs moving out take away density. Both these motions reduce the moment of inertia of the plasma and in the presence of any torques rotation shear can increase [35].

We postulate that a necessary condition for the L to H transition to take place is that the plasma background must have a magnetisation state boundary $\vec{j}_\theta = 0$ near the edge

$$\nabla p = \vec{j}_\zeta \times \vec{B}_\theta, \quad \text{i.e., } \vec{j}_\theta = 0 \tag{15}$$

At the edge of a steady L-mode plasma the toroidal current density can be estimated from loop voltage measurements and plasma resistivity. We therefore propose a testable necessary condition for H mode is that

$$|\nabla p| \geq |j_\zeta B_\theta| = \left| \frac{V_{loop}}{2\pi R \eta} B_\theta \right| \tag{16}$$

Here $\eta$ is the plasma resistivity. Such a condition could be checked against experimental measurements. We have not found these measurements in the published literature.

Further, if local measurements of poloidal current density in tokamaks become available, we could test unambiguously if a change in plasma background magnetisation is the trigger for transport barrier formation (L-H transition or formation of Internal Transport Barrier). As mentioned earlier, it is known that well developed transport barriers are associated with plasma diamagnetism [14].

Many other transport mechanisms are present in the plasma and in some cases compete with the magnetization instability. The merit of the simple model presented here is that it clearly singles out what we propose is the essential physical mechanism: magnetic phase separation.

**7. Summary:**

We show that considering tokamak plasmas as magnets that act on individual magnetised plasma blobs leads us to an understanding of magnetic phase transitions, and that they entail transport changes such as pedestal build up. A criterion to test the model is presented: it can hopefully be contrasted with experimental databases.

**Acknowledgements:**

We would like to thank Arturo Alonso (CIEMAT), P. Valanju (U. Texas), Dirk van Eester (ERM), Taina Kurki-Suonio (Aalto Univ.), G. Staebler (GA), B. Sieglin (IPP), J.A. Boedo (UCSD), Phil Edmonds and the referees for useful discussions, references and comments. One of the authors (ERS) is grateful to the Institute for Fusion Studies of the University of Texas at Austin for their hospitality during various visits, and to Rowena Ball and Bob Dewar of the Australian National University at Canberra, who provided partial funding for a visit in 2006 from Discovery grant DP0343765 and from COSNet grant RN0460006 of the Australian Research Council Complex Open Systems Research Network. The work was funded in part by the Contract of Association EURATOM-CIEMAT No. FU07-CT-2007-00050.